\documentclass[review,12pt]{elsarticle}
\usepackage{lineno,hyperref}
\modulolinenumbers[5]

\journal{Journal of \LaTeX\ Templates}

\usepackage{amsmath}
\usepackage{graphicx}

\title{ Stark-cyclotron Resonance in an Array of Carbon Nanotubes  }

\author[rvt]{S. S. Abukari\corref{cor1}}
\author[rvt]{S. Y. Mensah}
\author[els]{R. Musah}
\address[rvt]{Department of Physics, College of Agriculture and Natural Sciences, U.C.C, Ghana.}
\cortext[cor1]{Corresponding author\\ Email: sulemana70@gmail.com}
\ead[url]{sulemana70@gmail.com}
\author[focal]{ \\ N. G. Mensah}
\author[rvt]{ K. A. Dompreh}
\address[focal]{Department of Mathematics, College of Agriculture and Natural Sciences, U.C.C, Ghana}
\address[els]{Department of Applied Physics, University for Development Studies,
Navorongo, Ghana}

\date{}

\begin{document}
\begin{abstract}
\noindent Using the kinetic approach based on the semiclassical Boltzmann's transport equation with 
constant relaxation time, we theoretically studied the Stark-cyclotron resonance in an 
array of carbon nanotubes. Exact expression for the current density was obtained. 
We noted that Stark-cyclotron resonance occurs when the Larmor frequency coincides 
with the Stark frequency.

\end{abstract}

\maketitle
\section*{Introduction}
Carbon nanotubes (CNs) are allotropes of carbon with a nanometers in diameter and have a 
length-to-diameter ratio of the order $107$.  In $1952$, Radushkevich and Lukyanovich reported 
clear images of $50$ nanometer diameter tubes of carbon~\cite{1} and using the vapor-growth technique, 
Orberlin et al ~\cite{2} published hollow carbon fibers with nanometers in diameter. See ref~\cite{3} 
on additional reports on the observation of carbon nanotubes. However, the credit of the 
discovery of carbon nanotubes (CNs) goes to S. Iijima~\cite{4}. This one-atom thick sheet of 
graphene rolled up to a seamless cylinder with  a diameter of the order of a nanometer has 
since attracted a great deal of interest mainly due to their novel and unique thermal~\cite{5,6,7,8}, 
chemical and physical properties~\cite{9,10}. These properties depend on the fundamental indices ($n,m$) 
of the CNs. The indices ($n,m$)  determine the diameter and the chiral angle of the $CNs$. As $n$ and $m$ 
vary, the conduction ranges from metallic to semiconducting~\cite{11}, with an inverse diameter dependent 
band gap of $\leq 1eV$ ~\cite{11}. Electron transport properties in CNs have been the subject of intense 
research~\cite{10,11,12,13,14,15,16,17,18}. 
More recently, electronic transport in an array of CNs is the subject of many theoretical 
papers~\cite{19,20,21,22,23,24,25,26,27,28}.  Nevertheless, the electrodynamic properties of an array of 
CNs is worth further studying because it is the basis for developing carbon-based devices. 
Using the kinetic transport equation, we shall in this work study the effect of Stark cyclotron resonance 
in an array of CNs in the  presence of both constant electric and magnetic fields by following the 
approaches of~\cite{29,30}. 

\section*{Theory}
Proceeding as in references~\cite{29,30}, we consider the motion of an electron in the presence of both 
constant electric field $\vec{E}$ and magnetic field  $\vec{H}$. The electric field $\vec{E}$  is directed 
along the axis of the array of the CNs and  magnetic field  $\vec{H}$ directed at angle to the nanotubes axis. 
The CNs are arranged such that the distance between the neighboring CNs is larger than the CNs diameter 
and the interaction between the nanotubes is neglected~\cite{19,20,21,22,23,24,25,26,27,28}.  We shall 
take the CNs axes to be parallel to the $x$ –axis. The conductivity is derived using the Boltzmann kinetic 
equations describing electron transport in an array of CNs for the distribution functions in the relaxation 
time approximation as follows:
 \begin{equation}
e\vec{E} + \frac{e}{c}[V,\vec{H}]\frac{\partial f}{\partial {p}} = \frac{F - f}{\tau}
\end{equation}

where, $e$ is the electron charge.  The quasi-momentum is represented as ${p}=({p_x},s)$, where $p_x$
is the component of the electron dynamical momentum along the nanotube axis, while $s = 1,2...m$ is the number 
characterizing the quantization of momentum along the perimeter of the nanotube crosssection, $F$ and $f$ are 
the equilibrium and nonequilibrium distribution functions respectively. See~\cite{17,18,19,20,21,22,23,24,25,26,27}.
Taking into account the hexagonal crystalline structure of a unrolled graphene and using the tight binding approximation, 
the dispersion relation for the conduction electrons in the metallic zigzag CNs is given by~\cite{9, 17, 18}.
\begin{equation}
\varepsilon({p_x},s) = \pm\gamma_0 \sqrt{1 + 4cos(a{p_x})cos(\pi\frac{s}{m}) + 4cos^2(\pi\frac{s}{m})}
\end{equation}
where   $\gamma_0\approx 2.7 eV$,  $a=\frac{3b}{2\hbar}$, $b=0.142 nm$ is the distance between the neighbouring 
carbon atoms in the CNs. $+$ and $-$ signs are related to the conduction and valence bands respectively. 
The energy $\varepsilon({p_x},{p_y} )$  in Eq. ($1$) can be represented as a Fourier series~\cite{5, 13, 14}:
\begin{equation}
 \varepsilon({p_x},s) = \sum_{r =-\infty}^\infty{\varepsilon_{rs}exp i[r{p_x}]a}
\end{equation}
where $\hbar = 1$, and similarly, the expression for the distribution function in Fourier series are 
\begin{equation}
F({p_x},s) = \sum_{r =-\infty}^\infty{F_{rs} exp i[r{p_x}]a}
\end{equation}
where $F_{rs}$ and $\varepsilon_{rs}$. We define current density for the array of CNs are~\cite{19,20,21,22,23,24,25,26,27,28} as
\begin{equation}
j_x =\frac{e}{\tau}\sum_{s = 1}^{n}{\int_{-\frac{\pi}{a}}^{\frac{\pi}{a}}}F({p_x})d{p_x}\int_0^{\infty}
e^{-\frac{t}{\tau}}v({p_x},s)dt
\end{equation}
and the quasiclassical velocity $v(p_x )$  of an electron moving along the CNs axis ie $x$-component 
can be expressed as  
\begin{equation}
v(p_x,s)=\frac{\partial\varepsilon(p_x,s)}{\partial p_x} =i\gamma_0 a\sum_{r=-\infty}^{\infty}
r\varepsilon_{rs} exp i [rp_x]a
\end{equation}
The non relativistic equation motion of an electron in the presence of electric and magnetic fields 
which are constant in time and spatially uniform is given by ~\cite{29}
\begin{equation}
\frac{dp}{dt} = (e\vec{E} + \frac{e}{c} [v(p_x),\vec{H}])
\end{equation}
The components of Eq.($7$) are given by~\cite{28, 29, 30}
\begin{equation}
\frac{dp_x(t)}{dt} =e\vec{E} - \omega_{\perp}p_y(t)
\end{equation}
\begin{equation}
\frac{dp_y(t)}{dt} =\omega_{\perp}p_x(t) - \omega_{\parallel}p_z(t)
\end{equation} 
\begin{equation}
\frac{dp_z(t)}{dt} =\omega_{\parallel} p_y(t)
\end{equation}
where $\omega_{\parallel} = eBcos\theta/m^{\ast}$ are the parallel and perpendicular cyclotron frequencies. 
Asumming that $\vert \omega_{\parallel}\vert \gg \vert \omega_{\perp}\vert$, 
and solving Eqs.($8$)-($10$) we obtain
\begin{equation}
p_x(t) = p_x +\omega_B t -\frac{\omega_{\perp}}{\omega_{\parallel}}
p_{\perp} sin(\omega_{\parallel}t + \theta) + \frac{\omega_{\perp}}{\omega_{\parallel}}p_{\perp}sin(\theta)
\end{equation}
\begin{equation}
p_y(t) = p_{\perp}cos(\omega_{\parallel}t + \theta)
\end{equation}
\begin{equation}
p_z(t) = p_{\perp}sin(\omega_{\parallel}t + \theta)
\end{equation}
here $p_{\perp} =\sqrt{p_y^2 + p_z^2}$ and $tan{\theta} = p_z/p_y$ see~\cite{28,29,30}. 
Substituting Eq.($6$) into Eq.($5$) and using Eqn.($11$), we obtain the current density as
\begin{equation}
j_x =j_0\sum_{r=1}^{\infty}\sum_{m=-\infty}^{\infty}\sum_{s=1}^{n}{r\varepsilon_{rs}F_{rs}
j_m^2(\beta)[\frac{(r\Omega - k\omega_{\parallel})\tau}{1 + ((r\Omega - k\omega_{\parallel})\tau)^2}]}
\end{equation}
where $\Omega = eaE_{\parallel}$, $j_0 = 4e\gamma_0\sqrt{3}/\hbar^2$, $j_m $ is the Bessel function of 
$m^{th}$ order and $\beta =\frac{\omega_{\perp}3b}{\omega_{\parallel}\hbar}p_{\perp}$. 

\begin{figure}
\begin{centering}
\includegraphics[width = 12cm]{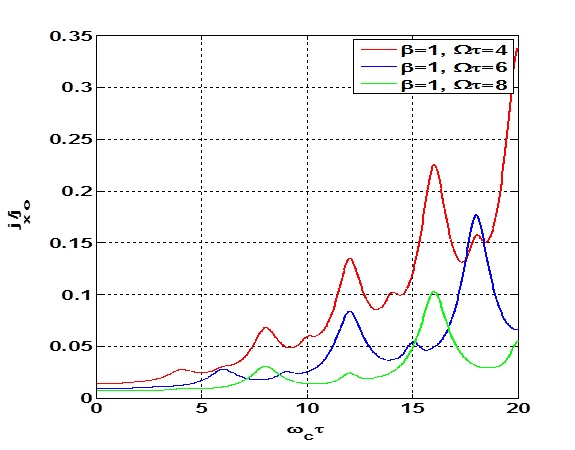} 
 \caption{ A plot of a normalized current density $j_x/j_0$    as a function of  $\omega_c\tau$    
in a An Array of CNs for expression ($8$) when $\Omega\tau= 5, 10$, and $15...$} 
\end{centering}
\end{figure}
\section*{Results, Discussion and Conclusion}
We present the results of a kinetic equation approach of a $2D$ array of zigzag CNs 
subject to both constant electric field $\vec{E}$ and magnetic field  $\vec{H}$. 
Exact expression for the direct current density was 
obtained in eq. ($14$). The nonlinearity is analyzed using the dependence of the 
normalized direct current density $j_x/j_0$  as a function of $\omega_c\tau$. A plot 
of a normalized current density $j_x/j_0$    as a function of $\omega_c\tau$   in 
metallic zigzag CNs array for expression ($8$) when $\Omega\tau= 4,6$ and $8$ is shown 
in Fig. $1$. We observed that resonance occurs when the ratio of the Stark and the 
cyclotron frequencies is an integer (i.e $k\omega_c\tau = \Omega\tau$). The peak current 
density corresponding to the Stark-cyclotron resonance also shifts towards large $\omega_c\tau$   
values with increasing $\Omega\tau$ values. 
In conclusion, we considered the nonlinear properties in impure graphene subject to electric 
and magnetic fields. The results indicate Stark-cyclotron resonance which is in essence due 
to nonparabolicity of the energy spectrum of the CNs.

\renewcommand\refname{Bibliography}

\end{document}